\documentclass[12pt,a4paper]{article}
\usepackage{amsmath,amssymb}

\usepackage{ifpdf}
\ifpdf
  \usepackage[pdftex]{graphicx}
\else
  \usepackage[dvipdfm]{graphicx}
\fi

\newsavebox{\boxa}

\begin{document}

\thispagestyle{empty}
\begin{flushright}
KEK-TH 1488
\end{flushright}
\vskip3cm
\begin{center}
{\Large {\bf A note on statistical model for BPS D4-D2-D0
\\[2mm]
 states}}
\vskip1.5cm
{\large 
{Takahiro Nishinaka\footnote{nishinak [at] post.kek.jp}
}
and\hspace{2mm} Yutaka Yoshida\footnote{yyoshida [at] post.kek.jp}
}
\vskip.5cm
{\it High Energy Accelerator Research Organization (KEK),
\\
Tsukuba, Ibaraki 305-0801, Japan}
\end{center}

\vskip2cm
\begin{abstract}
 We construct a statistical model that reproduces the BPS partition function of D4-D2-D0 bound states on a class of toric Calabi-Yau three-folds. The Calabi-Yau three-folds we consider are obtained by adding a compact two-cycle to $A_{N-1}$-ALE $\times\; \mathbb{C}$. We show that in the small radii limit of the Calabi-Yau the D4-D2-D0 partition function is correctly reproduced by counting the number of triangles and parallelograms.
\end{abstract}


\newpage

It is a longstanding problem in string theory to count the BPS bound states of D-branes wrapping on a Calabi-Yau three-fold. For a compact Calabi-Yau three-fold the wrapped D-branes can be seen as BPS black holes in four-dimensional supergravity, while if the Calabi-Yau is non-compact then the D-branes correspond to BPS particles in four-dimensional supersymmetric gauge theory. Therefore, such a counting problem is relevant for the entropy of BPS black holes and the spectrum of BPS particles in four dimensions.

Recently, there has been remarkable progress in the study of the D-brane counting, especially for D6-D2-D0 and D4-D2-D0 states on a toric Calabi-Yau three-fold. In particular, one of the most interesting results is that the stable BPS bound states of D-branes have some statistical model descriptions. For D6-D2-D0 states, it is known that the crystal melting model correctly describes the BPS bound states of D6-D2-D0 branes on $\mathbb{C}^3$ \cite{Okounkov:2003sp}, resolved conifold \cite{Szendroi:2007nu} (See also \cite{Chuang:2008aw}), and general toric Calabi-Yau three-folds \cite{Mozgovoy:2008fd, Ooguri:2008yb}. Their D4-D2-D0 counterpart was studied in \cite{Nishinaka:2011sv} only on the resolved conifold.\footnote{For another interesting work on the statistical model description of D4-D2-D0 states, see \cite{Szabo:2011mj}.} The statistical model description of the BPS D-branes provided some insights on the quantum description of geometry in string theory \cite{Iqbal:2003ds, Ooguri:2009ri}.

\begin{figure}
\begin{center}
\includegraphics[width=7cm]{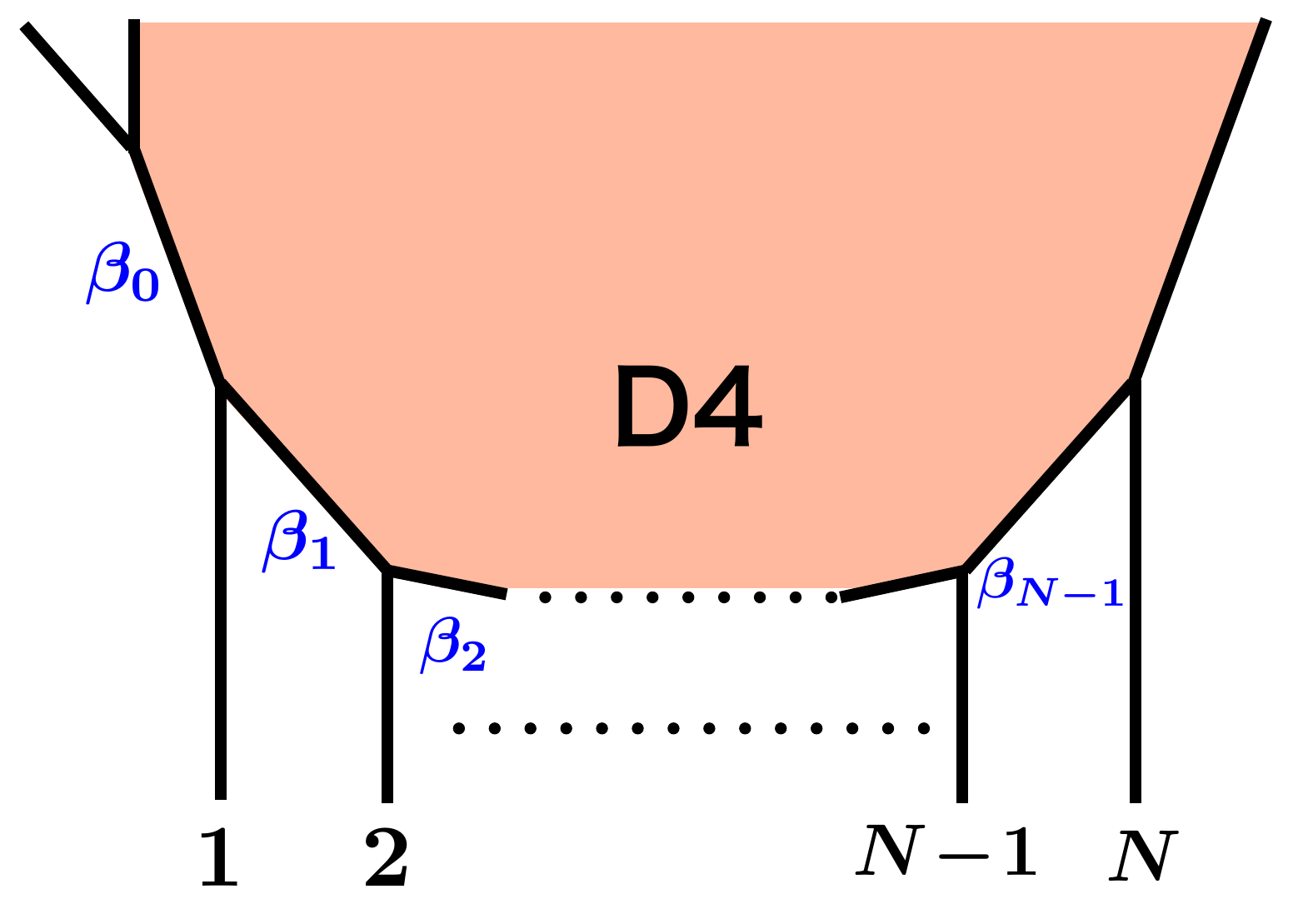}
\caption{The toric webdiagram of the Calabi-Yau three-fold we consider. It can be obtained by adding a compact two-cycle $\beta_0$ to $A_{N-1}$-ALE $\times\;\mathbb{C}$. In the vicinity of the two-cycle $\beta_0$, the Calabi-Yau can be seen as the resolved conifold $\mathcal{O}(-1) \oplus \mathcal{O}(-1)\to\mathbb{P}^1$.}\label{fig:diagram}
\end{center}
\end{figure}

In this article, we generalize the statistical model description of BPS D4-D2-D0 states studied in \cite{Nishinaka:2011sv} to a more complicated toric Calabi-Yau three-fold. The toric Calabi-Yau three-fold $X$ we consider is obtained by adding a compact two-cycle to $A_{N-1}$-ALE $\times\;\mathbb{C}$. Its webdiagram can be depicted as in figure \ref{fig:diagram}.
 Such a Calabi-Yau three-fold was considered in \cite{Nishinaka:2011nn} to study the relation between the instanton partition functions on ALE space and $\mathbb{C}^2$. We put a non-compact D4-brane on a divisor of $X$, which corresponds to the shaded region of figure \ref{fig:diagram}, and consider the BPS bound states of D0-branes localized in $X$ and D2-branes wrapped on compact two-cycles in $X$.

The BPS partition function of interest is defined by
\begin{eqnarray}
 \mathcal{Z}(q,Q) &:=& \!\!\!\!\sum_{n,m^0,m^1,\cdots,m^{N-1}}\!\!\!\!\Omega(\mathcal{D} + m^I\beta_I - ndV)\; q^n\, Q_0^{m^0}Q_1^{m^1}\cdots Q_{N-1}^{m^{N-1}},
\label{eq:def_partition_function}
\end{eqnarray}
where $\Omega(\Gamma)$ is the BPS index of charge $\Gamma$, and $I$ runs over $I=0,1,2,\cdots,N-1$. The two-form $\mathcal{D}\in H^{2}(X)$ denotes the charge for a single non-compact D4-brane wrapped on the divisor of $X$. The four-form $\beta_I\in H^4(X)$ represents the unit charge of D2-brane wrapped on $I$-th two-cycle of the Calabi-Yau $X$. The unit D0-brane charge is denoted by $-dV$.\footnote{In this letter, we use the same notation as in \cite{Nishinaka:2011nn}. In particular, we have $\int_X dV=1$.} We denote by $q$ and $Q_I$ the Boltzmann weights for D0-branes and D2-branes on $I$-th cycle, respectively.

As shown in \cite{Nishinaka:2011nn}, the partition function \eqref{eq:def_partition_function} has a non-trivial moduli dependence. Since we are considering type IIA string theory on a Calabi-Yau three-fold, the BPS partition function depends on the K\"ahler moduli of the Calabi-Yau $X$, which are roughly the sizes and B-fields of the compact two-cycles in $X$. We denote by $z_I$ the complexified K\"ahler parameter for the $I$-th compact two-cycle, so that a D2-brane wrapped on the $I$-th two-cycle has the central charge $z_I$.
Then, it was shown in \cite{Nishinaka:2011nn} that if we tune the moduli parameter $z_I$ so that ${\rm Im}\,z_I = 0,\;0<{\rm Re}\,z_I<1$ and $0 < {\rm Re}\,z_0 < 1-\sum_{i=1}^{N-1}{\rm Re}\,z_i$ then the BPS partition function is given by\footnote{Although this is not explicitly stressed in \cite{Nishinaka:2011nn}, it is easily seen as follows. In \cite{Nishinaka:2011nn}, it was pointed out that the partition function is given by $\mathcal{Z} = \prod_{m=1}^\infty(1-q^n)^{-1}$ when ${\rm Im}\,z_0 = -\infty$ and $0<{\rm Re}\,z_0 <1 - \sum_{i=1}^{N-1}{\rm Re}\,z_i$. By moving ${\rm Im}\,z_0$ from ${\rm Im}\,z_0 = -\infty$ to ${\rm Im}\,z_0 = 0$, various walls of marginal stability are crossed, which gives rise to discontinuous changes in the partition function. From figure 12 of \cite{Nishinaka:2011nn}, we find that such discontinuous changes between ${\rm Im}\,z_0 =-\infty$ and ${\rm Im}\,z_0=0$ can be written as equation (4.21) in \cite{Nishinaka:2011nn} when we keep ${\rm Im}\,z_1=\cdots = {\rm Im}\,z_{N-1}=0$. Then, the resulting partition function at ${\rm Im}\,z_0 = 0$ turns out to be written as equation \eqref{eq:partition_function} in this article.}
\begin{eqnarray}
\mathcal{Z}(q,Q) &=& \prod_{m=1}^\infty\frac{1}{1-q^m}\prod_{n=0}^\infty(1-q^nQ_0)(1-q^nQ_0Q_1)\cdots(1-q^nQ_0Q_1\cdots Q_{N-1}).
\label{eq:partition_function}
\end{eqnarray}
Below we construct a two-dimensional {\em oblique partition model} whose partition function is exactly the same as the above D4-D2-D0 partition function \eqref{eq:partition_function}.

\subsubsection*{Oblique partition model}

\begin{figure}
\begin{center}
\includegraphics[width=7.5cm]{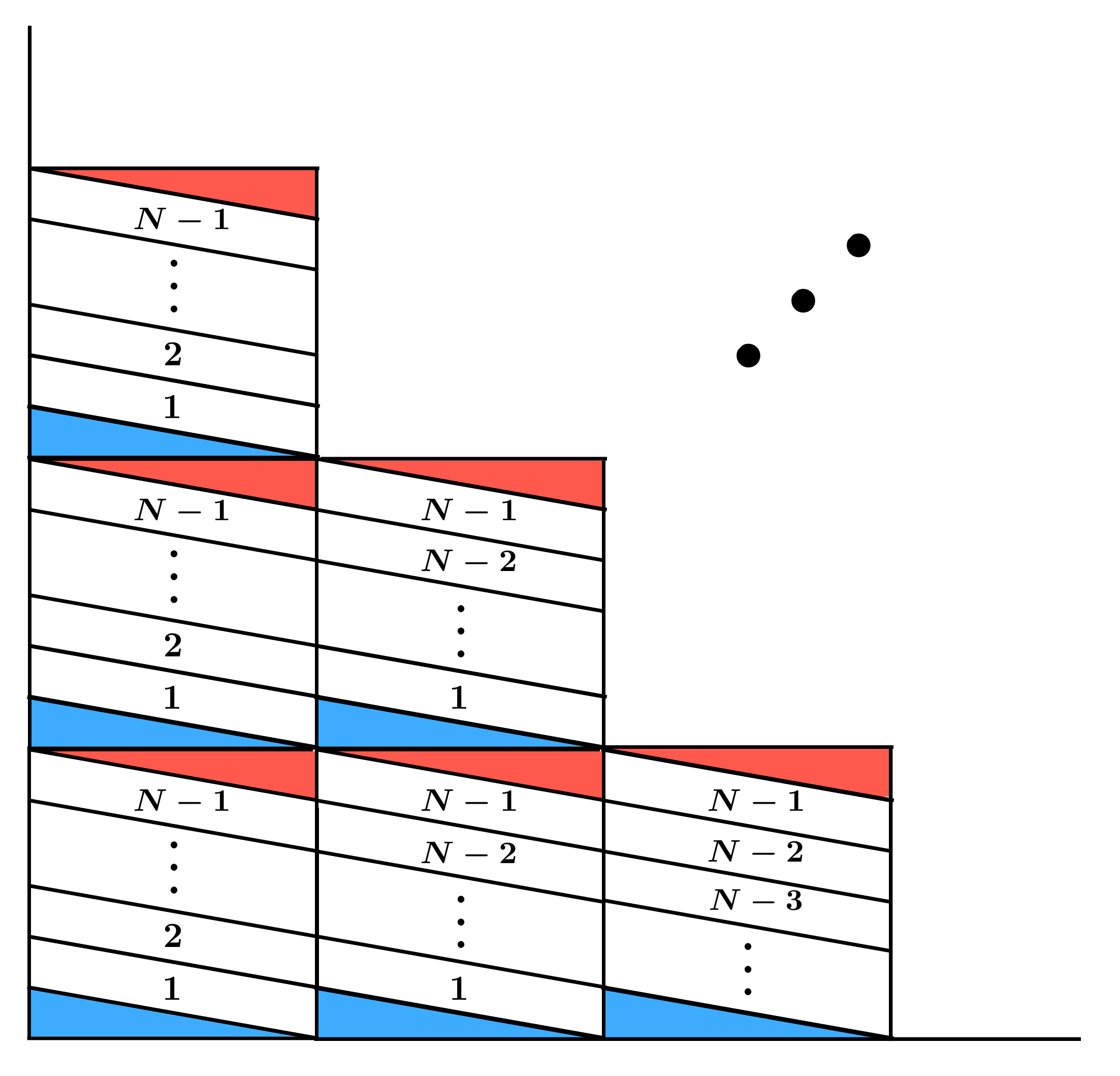}
\qquad\qquad
\includegraphics[width=3cm]{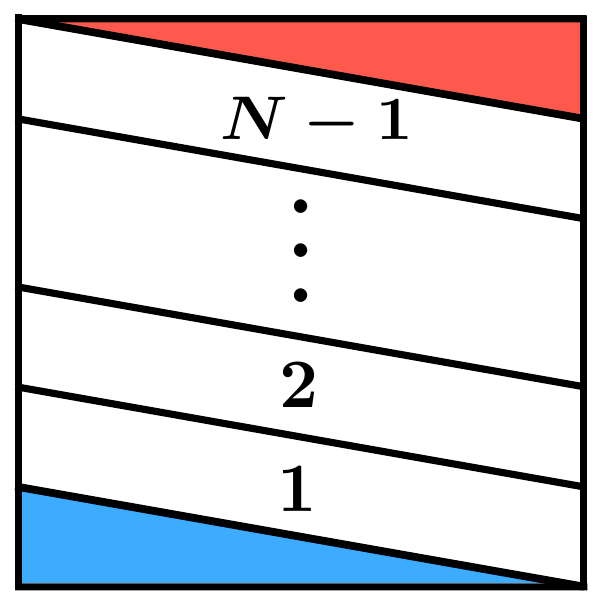}
\caption{Left: The two-dimensional crystal composed of $N+1$ kinds of atoms, that is, two kinds of triangle atoms and $N-1$ kinds of parallelogram atoms. The crystal fills the upper right plane. \; Right: The building block of the crystal. The blue and red triangles are at the bottom and top, respectively. In the middle of the building block, there are $N-1$ numbered parallelograms.}
\label{fig:crystal}
\end{center}
\end{figure}

The oblique partition model which we propose is the following. First of all, let us consider a two-dimensional crystal that is composed of $N+1$ kinds of atoms, as in the left picture of figure \ref{fig:crystal}. 
More precisely, it contains two kinds of triangle atoms and $N-1$ kinds of parallelogram atoms. The former is classified by their colors (blue or red), while the latter is assigned the numbers $1,2,\cdots, N-1$. The crystal is infinitely extended in the upper right plane. The building block of the crystal is depicted in the right picture of figure \ref{fig:crystal}. The blue and red triangles are at the bottom and top of the building block, respectively, while in the middle there are $N-1$ numbered parallelograms. We associate the whole crystal to a D4-brane without any D2 and D0-brane charges. From \eqref{eq:partition_function}, it follows that there is only one such state.

\begin{figure}
\begin{center}
\includegraphics[width=7cm]{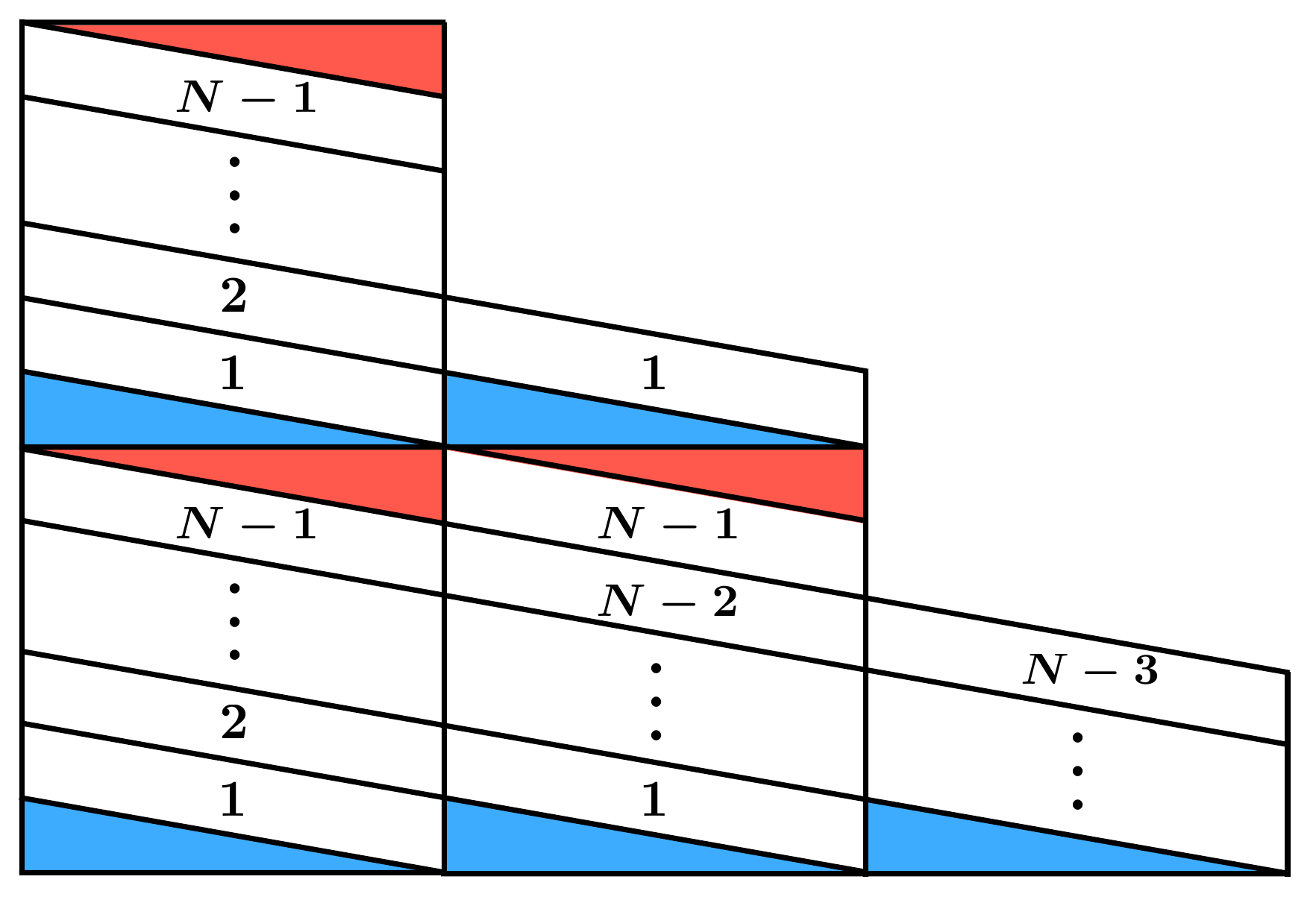}
\qquad
\includegraphics[width=7cm]{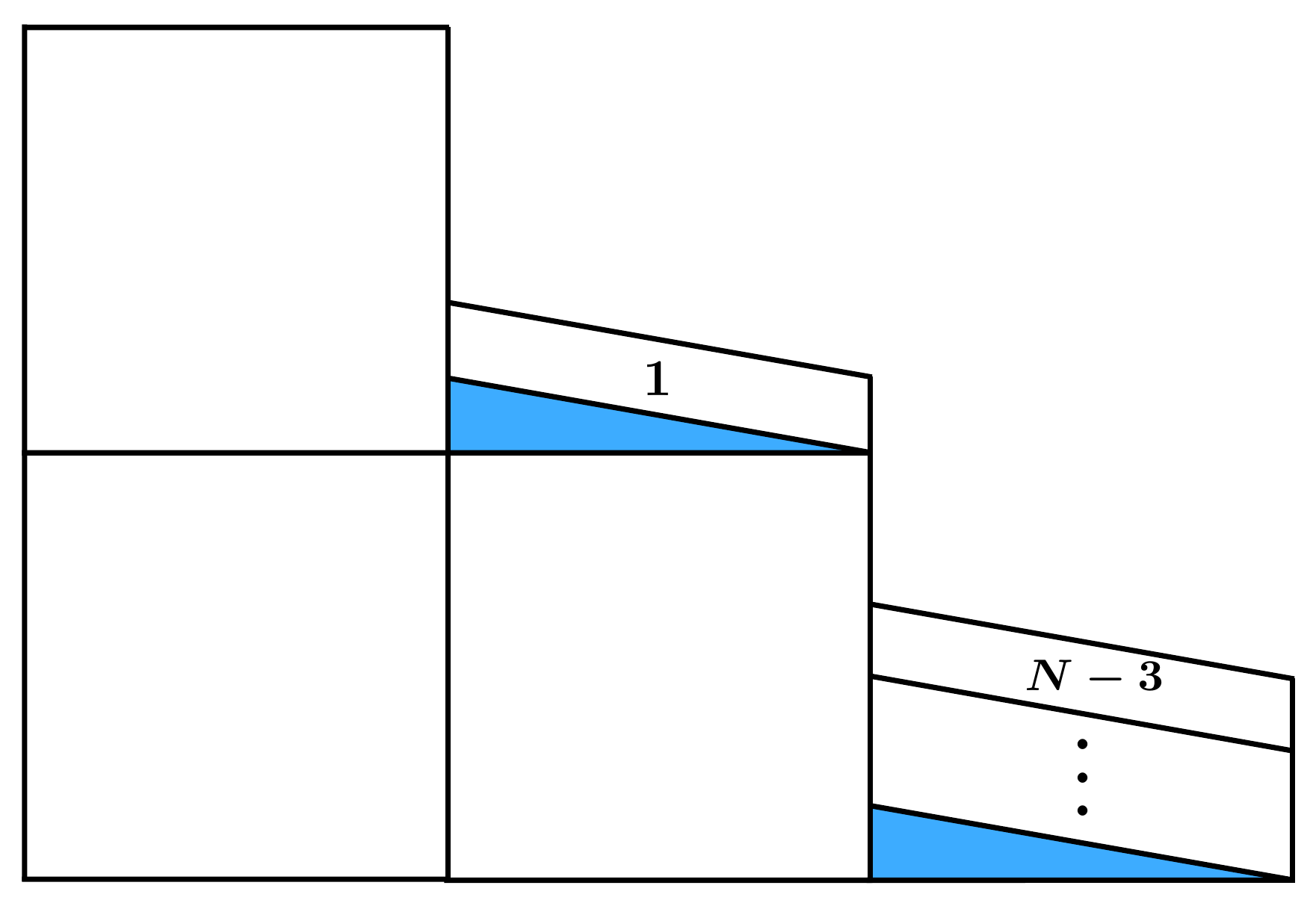}
\caption{Left: A typical example of ``the oblique partition'' (a set of atoms that can be removed from the crystal).\; Right: A set of a blue triangle, a red triangle and $N-1$ different parallelograms forms a square building block of figure \ref{fig:crystal}, which should be counted as a single D0-brane. The number of the remaining blue triangle is identified with the zero-th D2-brane charge $m^0$ (up to the sign attached to the Boltzmann weight), while the number of the remaining $i$-th parallelograms is regarded as the $i$-th D2-brane charge $m^i$ for $i=1,2,\cdots,N-1$. The above example has $n=3,\,m^0=m^1=2,\,m^{2}=\cdots=m^{N-3}=1$ and $m^{N-2}=m^{N-1}=0$.}
\label{fig:oblique_partition}
\end{center}
\end{figure}

We now remove some of the atoms from this crystal under the following rules:
\begin{enumerate}
\item A blue triangle can be removed only if two of its sides are not adjoined to other atoms.
\item A red triangle can be removed only if one of its sides is not adjoined to other atoms.
\item A parallelogram can be removed only if two of its sides are not adjoined to other atoms.
\end{enumerate}
Let us call a set of atoms that can be removed from the crystal ``an oblique partition.'' A typical example of the oblique partition is depicted in the left picture of figure \ref{fig:oblique_partition}.
Then, let us define $a,b$ and $c_i$ for an oblique partition by
\begin{eqnarray}
 a &=& (\mbox{the number of red triangles in the oblique partition}),
\nonumber \\
 b &=& (\mbox{the number of blue triangles in the oblique partition}),
\nonumber \\
c_i &=& (\mbox{the number of }i\mbox{-th parallelograms in the oblique partition}).
\nonumber 
\end{eqnarray}
From the above rules of removing atoms, it always follows that
\begin{eqnarray}
b\geq c_1\geq c_2\geq \cdots\geq c_{N-1}\geq a.
\label{eq:inequality}
\end{eqnarray}
We define the partition function of this oblique partition model by
\begin{eqnarray}
 \mathcal{Z}_{\rm oblique} &=& \sum_{\rm oblique\;partitions}\left((-1)^{a+b}x^a y^b \prod_{i=1}^{N-1}z_i^{c_i}\right),
\end{eqnarray}
where $x,y$ and $z_i$ are the Boltzmann weights for red triangles, blue triangles, and $i$-th parallelograms, respectively.

We have already associated the crystal without removing atoms to a D4-brane without D2 and D0-brane charges. Now, we associate each oblique partition to a D4-D2-D0 bound state, by relating $a,b$ and $c_i$ to the D2 and D0-brane charges. We first relate the Boltzmann weights for D0 and D2-branes to those for atoms by
\begin{eqnarray}
 q = xy\prod_{i=1}^{N-1}z_i,\qquad Q_0 = y, \qquad Q_i = z_i \quad {\rm for}\quad i=1,2,\cdots,N-1.
\label{eq:relation}
\end{eqnarray}
Recall here that we have $N$ independent D2-brane charges, which are denoted by $m^I$ for $I=0,1,2,\cdots,N-1$. The D0-brane charge is expressed by $n$. Hence, the above relation \eqref{eq:relation} implies that
\begin{eqnarray}
 n = a, \qquad m^0 = b-a,\qquad m^i = c_i-a\quad {\rm for}\quad i=1,2,\cdots,N-1.
\end{eqnarray}

The relation \eqref{eq:relation} implies that the D0-brane charge $n$ can be regarded as the number of building blocks that are included in the oblique partition. A set of a blue triangle, a red triangle and $N-1$ different parallelograms forms a square building block of the right picture of figure \ref{fig:crystal}, which should be counted as a single D0-brane (See the right picture of figure \ref{fig:oblique_partition}). Since we have the inequality \eqref{eq:inequality} for every oblique partition, the second relation of \eqref{eq:relation} implies that the numbers of the remaining blue triangles should be identified with the zero-th D2-brane charge $m^0$. Similarly, the $i$-th D2-brane charge $m^i$ for $i=1,2,\cdots,N-1$ is identified with the number of the remaining $i$-th parallelograms. Note that we can always form the square building blocks so that all the remaining blue triangles and parallelograms are left at the upper edge of the oblique partition.

\subsubsection*{Reproducing the D4-D2-D0 partition function}

We now show that with the identification \eqref{eq:relation} the oblique partition model precisely reproduces the D4-D2-D0 partition function \eqref{eq:partition_function}, that is, 
\begin{eqnarray}
 \mathcal{Z}_{\rm oblique} &:=& \sum_{\rm oblique\;partitions}\left((-1)^{a+b}x^ay^b \prod_{i=1}^{N-1}z_i^{c_i}\right)
\nonumber \\[2mm]
&=&  \prod_{m=1}^\infty\frac{1}{1-q^m}\prod_{n=0}^\infty(1-q^nQ_0)(1-q^nQ_0Q_1)\cdots(1-q^nQ_0Q_1\cdots Q_{N-1}).
\label{eq:equivalence}
\end{eqnarray}
In order to make our idea clear, we first concentrate on the simplest case $N=2$, and will later generalize it to the case of $N\geq 3$.

\begin{figure}
\begin{center}
\includegraphics[width=6cm]{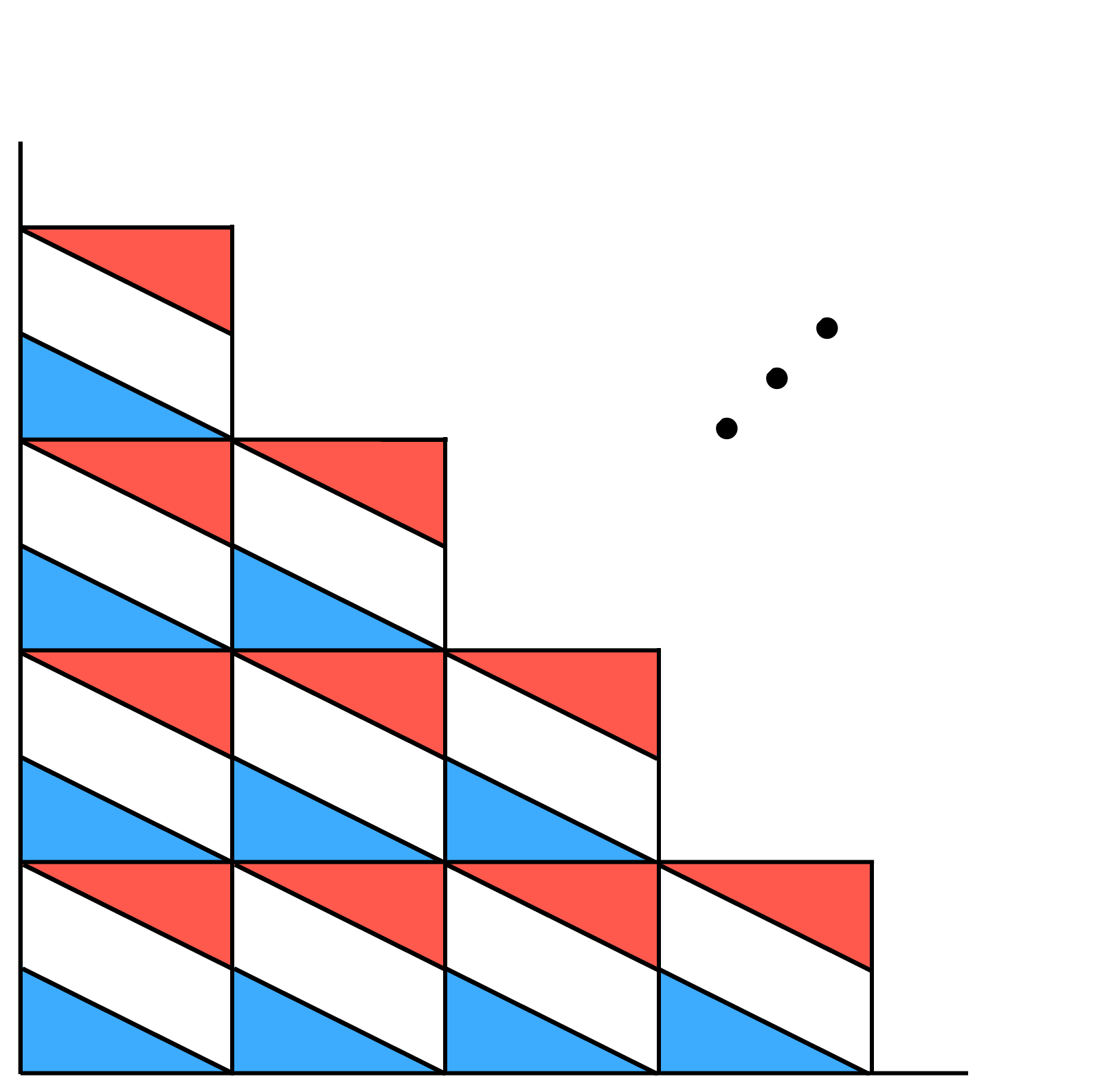}
\qquad\qquad
\includegraphics[width=5cm]{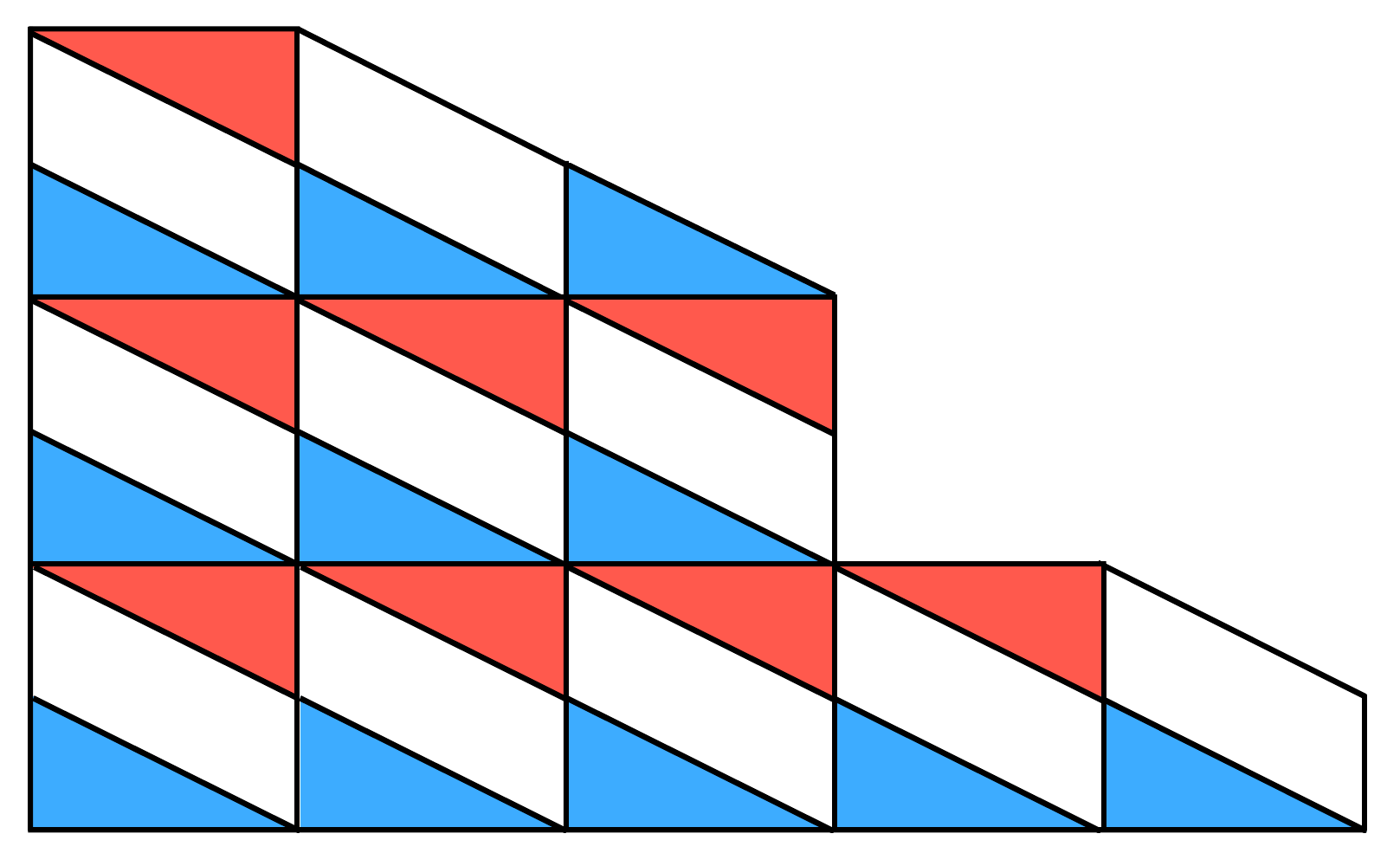}
\caption{Left: The crystal in the case of $N=2$. There are three kinds of atoms, that is, blue triangles, red triangles and white parallelograms. Since there is only a single type of parallelogram, we omit to write the numbers in the parallelograms.\; Right: A typical ``oblique partition'' (the set of atoms that can be removed from the crystal under our rules).}
\label{fig:N=2}
\end{center}
\end{figure}

\begin{figure}
\begin{center}
\includegraphics[width=5cm]{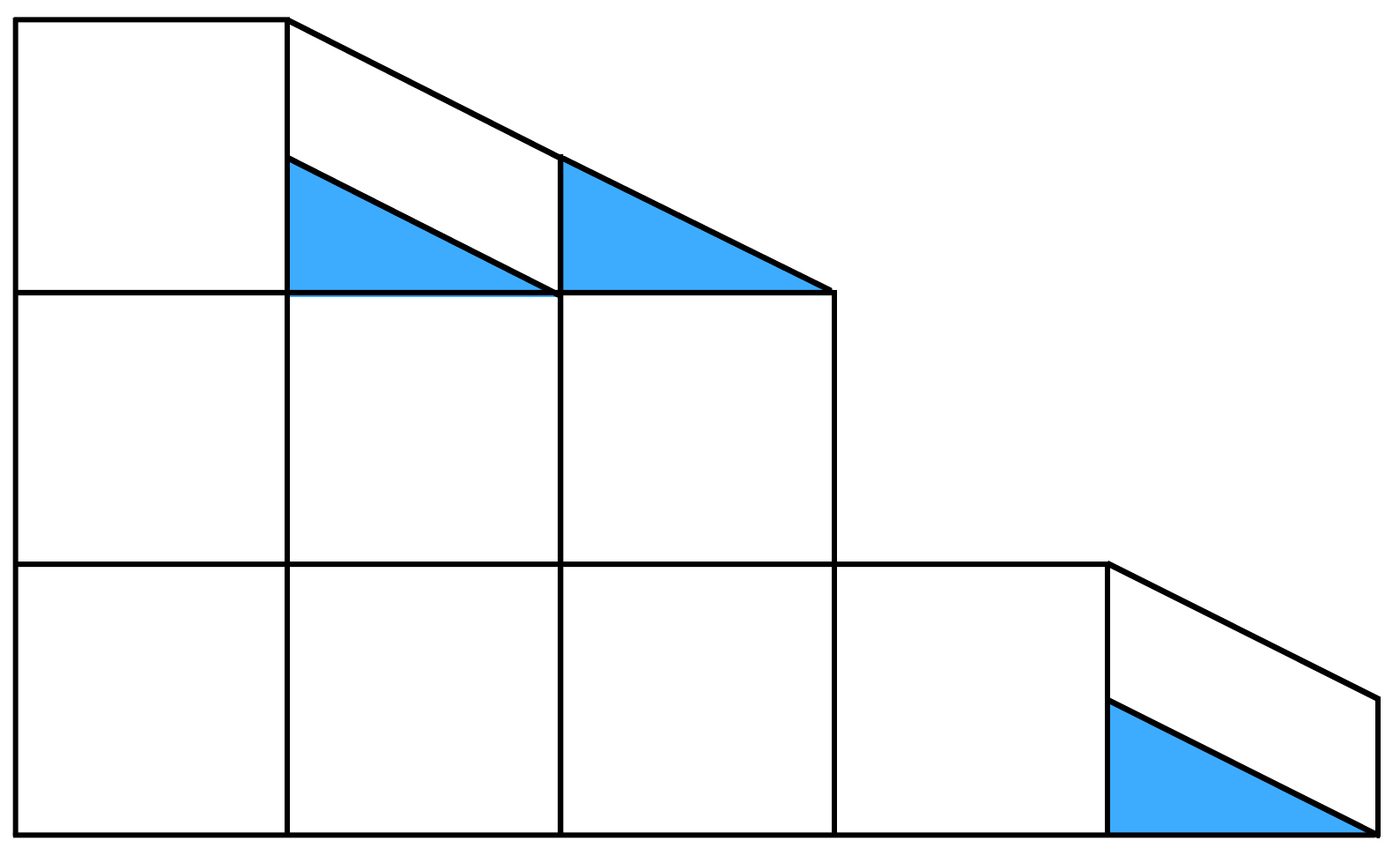}
\qquad\qquad
\includegraphics[width=6cm]{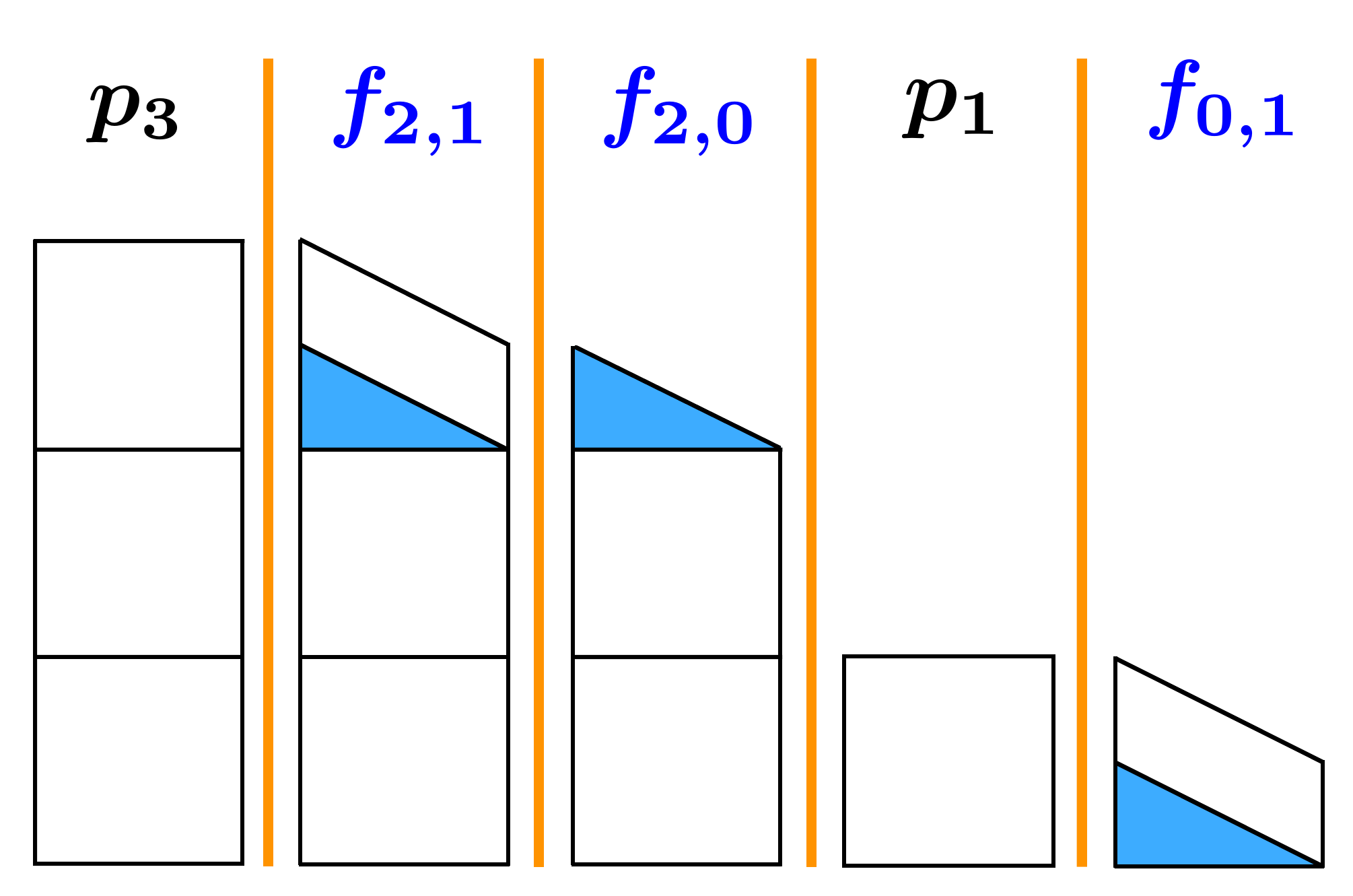}
\caption{Left: By treating a square building block as a single constituent, all the red triangles are absorbed into the white squares. \; Right: We cut up the oblique partition into the towers of the squares. Some of the tower have a parallelogram and/or a blue triangle on their tops, while the others only contain square building blocks. Collecting the latter gives a two-dimensional Young diagram.}
\label{fig:divide_atoms:N=2}
\end{center}
\end{figure}

In the case of $N=2$, the whole crystal and a typical oblique partition can be depicted as in figure \ref{fig:N=2}, where we omit to write the numbers in the parallelograms because there is only a single type of the parallelogram for $N=2$. 
By treating a square building block as a single constituent, all the red triangles in the oblique partition are absorbed into the building blocks (figure \ref{fig:divide_atoms:N=2}). The number of such squares is identified with the D0-brane charge. 

\begin{figure}
\begin{center}
\includegraphics[width=5.5cm]{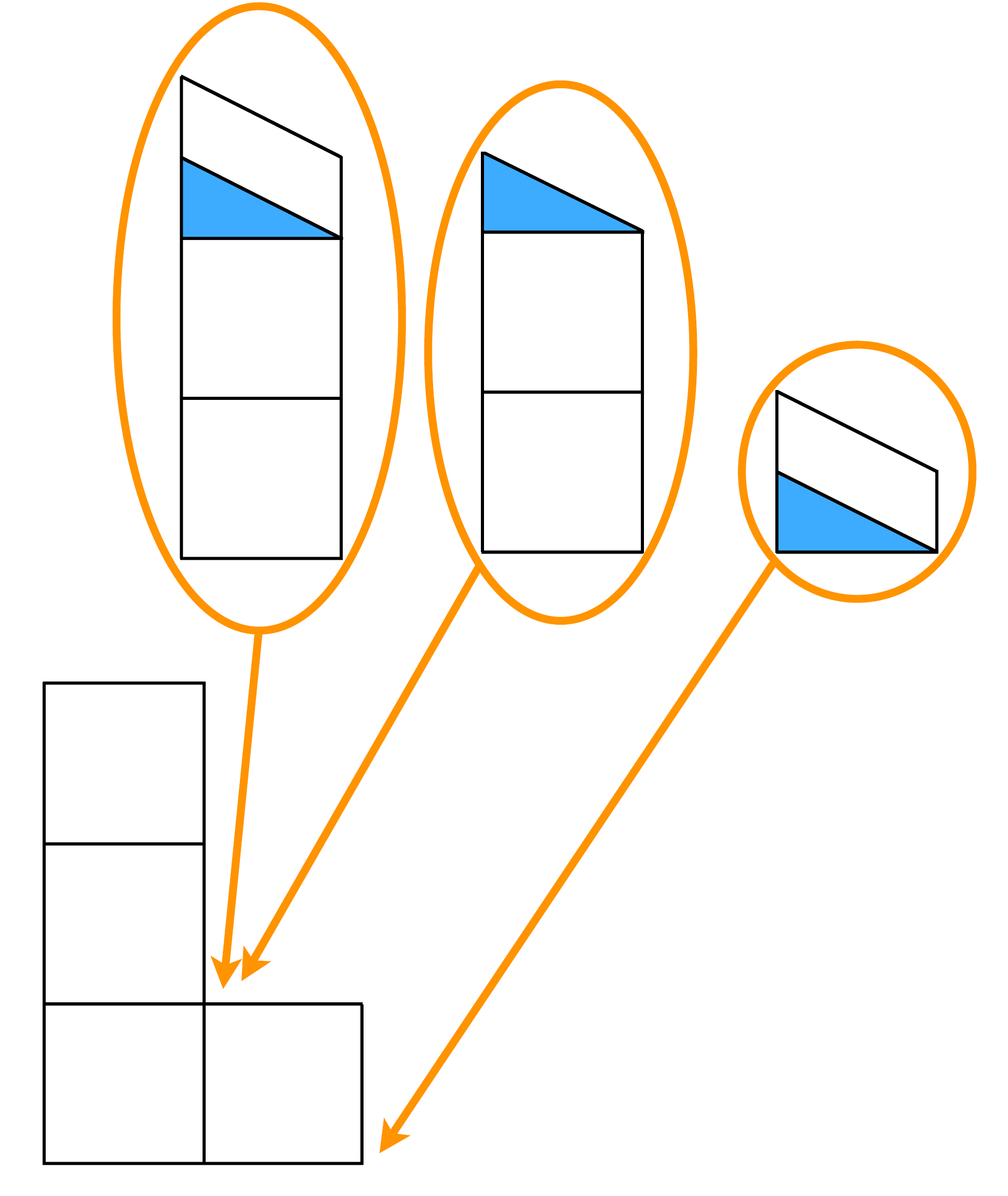}
\caption{An oblique partition can be reconstructed by inserting some towers of $f_{n,k}$ into a Young diagram.}
\label{fig:insertion}
\end{center}
\end{figure}

To prove the equivalence \eqref{eq:equivalence}, we first cut up the oblique partition into the towers of square building blocks, as in the right picture of figure \ref{fig:divide_atoms:N=2}. 
Some of the towers have a parallelogram and/or a blue triangle on their tops, while the others only contain square building blocks. We denote by $p_n$ a tower of the latter that contains $n$ square building blocks (See figure \ref{fig:divide_atoms:N=2}). Collecting them gives a usual two-dimensional Young diagram.
On the other hand, the former is labeled by the number of squares $n$ and one additional parameter $k$.
We set $k=0$ if the tower contains a blue triangle without any other parallelograms. If the tower includes some parallelograms, then we let $k$ be the number of such parallelograms. In the case of $N=2$, we have only two cases; $k=0$ or $k=1$. We denote each such tower by $f_{n,k}$ (See the right picture of figure \ref{fig:divide_atoms:N=2}).
In other words, an oblique partition is reconstructed by inserting some towers of $f_{n,k}$ into a Young diagram, as in figure \ref{fig:insertion}.

Our rules of removing atoms implies that the insertion of $f_{n,k}$ for fixed $(n,k)$ cannot be made more than once. So these insertions of $\{f_{n,k}\}$ are ``fermionic.''  Moreover, for an arbitrary $f_{n,k}$, the position of the insertion is uniquely determined so that the resulting oblique partition can be removed from the crystal under our rules. To be more precise, the insertion of $f_{n,k}$ should be made so that the following conditions are met:
\begin{enumerate}
\item All towers of $p_m$ for $m\leq n$ exist in the right side of the insertion, while those for $n<m$ exist in its left.
\item All towers of $f_{m,l}$ for $m<n$ exist in the right side of the insertion, while those for $n<m$ exist in its left.
\item All towers of $f_{n,l}$ for $l<k$ exist in the right side of the insertion, while those for $k<l$ exist in its left.
\end{enumerate}
Note that for any $f_{n,k}$ there exists a single insertion locus satisfying these conditions.
The above argument implies that {\em the oblique partitions are in one-to-one correspondence with the fermionic insertions of $\{f_{n,k}\}$ into two-dimensional Young diagrams.}

According to the identification rules \eqref{eq:relation}, we find that the insertion of $f_{n,0}$ contributes $-q^n Q_0$ to the partition function, while the insertion of $f_{n,1}$ contributes $-q^nQ_0Q_1$. Therefore, all the fermionic insertions of $\{f_{n,k}\}$ for $k=0,1$ are taken into account by multiplying 
\begin{eqnarray}
 \prod_{n=0}^{\infty}(1-q^n Q_0)(1-q^n Q_0Q_1).
\end{eqnarray}
On the other hand, summing up all the two-dimensional Young diagrams gives
\begin{eqnarray}
 \prod_{m=1}^{\infty}\frac{1}{1-q^m}.
\label{eq:Young}
\end{eqnarray}
Hence, {\em by collecting all the fermionic insertions of $\{f_{n,k}\}$ into all the Young diagrams, we obtain the collect expression of the D4-D2-D0 partition function \eqref{eq:partition_function}, which gives the proof of the equivalence \eqref{eq:equivalence} in the case of $N=2$.} This implies that our oblique partition model correctly reproduces the BPS partition function of D4-D2-D0 states.

Now, we generalize the above arguments to the case of $N\geq 3$. In fact, such a generalization is straightforward. Most of the above arguments are also applicable to a general case. The only one difference from the simplest case of $N=2$ is that we now have three or more kinds of fermionic insertions $f_{n,k}$ for each value of $n$. The integer $k$ of $f_{n,k}$ generally has $N$ possible values of $k=0,1,2,\cdots,N-1$.\footnote{Recall that the integer $k$ of $f_{n,k}$ is the number of the parallelograms on the top of the tower of square building blocks. In general, there are $N-1$ different types of parallelograms in the crystal.} However, even in such a case, it is still true that the oblique partitions are in one-to-one correspondence with the fermionic insertions of $\{f_{n,k}\}$ into Young diagrams. Now, the insertion of $f_{n,k}$ contributes $-q^nQ_0\cdots Q_{k}$ to the partition function. So all such insertions are taken into account by multiplying
\begin{eqnarray}
 \prod_{n=0}^\infty (1-q^nQ_0)(1-q^nQ_0Q_1)\cdots (1-q^nQ_0\cdots Q_{N-1}),
\end{eqnarray}
to \eqref{eq:Young}, which correctly reproduces the right-hand side of \eqref{eq:partition_function}. This gives a proof of the equivalence \eqref{eq:equivalence} for a general value of $N$.

\subsubsection*{Relation to fractional branes}

We now briefly mention the relation to the fractional D0-branes.\footnote{The authors thank Yoshifumi Hyakutake, Muneto Nitta and Kazutoshi Ohta for pointing out the possible relation between the statistical model description of D4-D2-D0 states and the fractional branes.} In our statistical model,  the $i$-th  prallelogram is associated with the charge of D2-brane wrapping $i$-th
cycle $\beta_i, \, (1 \le i \le N-1)$ and the  blue triangle    is associated with the charge of D2-brane wrapping on 
the  cycle $\beta_0$.
On the other hand, unit D0-brane charge is labeled by a single square box which 
can be also regarded as the cluster of $N-1$ kinds of prallelograms  
and two kinds of  triangles (figure \ref{fig:crystal}). This feature of the model 
implies that the union of $N$ kinds of D2-branes (and a red triangle) can be interpreted as  D0-branes with 
unit charge and remind us   the relationship between  D2-branes and fractional D0-branes on the $A_{N-1}$-type ALE space.
In the orbifold limit of the ALE space, a single D2-brane wrapping  the vanishing  two-cycle is described by a D0-brane localized at the orbifold point with fractional charge \cite{Diaconescu:1997br}, and the union of $N$ such fractional D0-branes can be a unit regular D0-brane.

\begin{figure}
\begin{center}
\includegraphics[width=8cm]{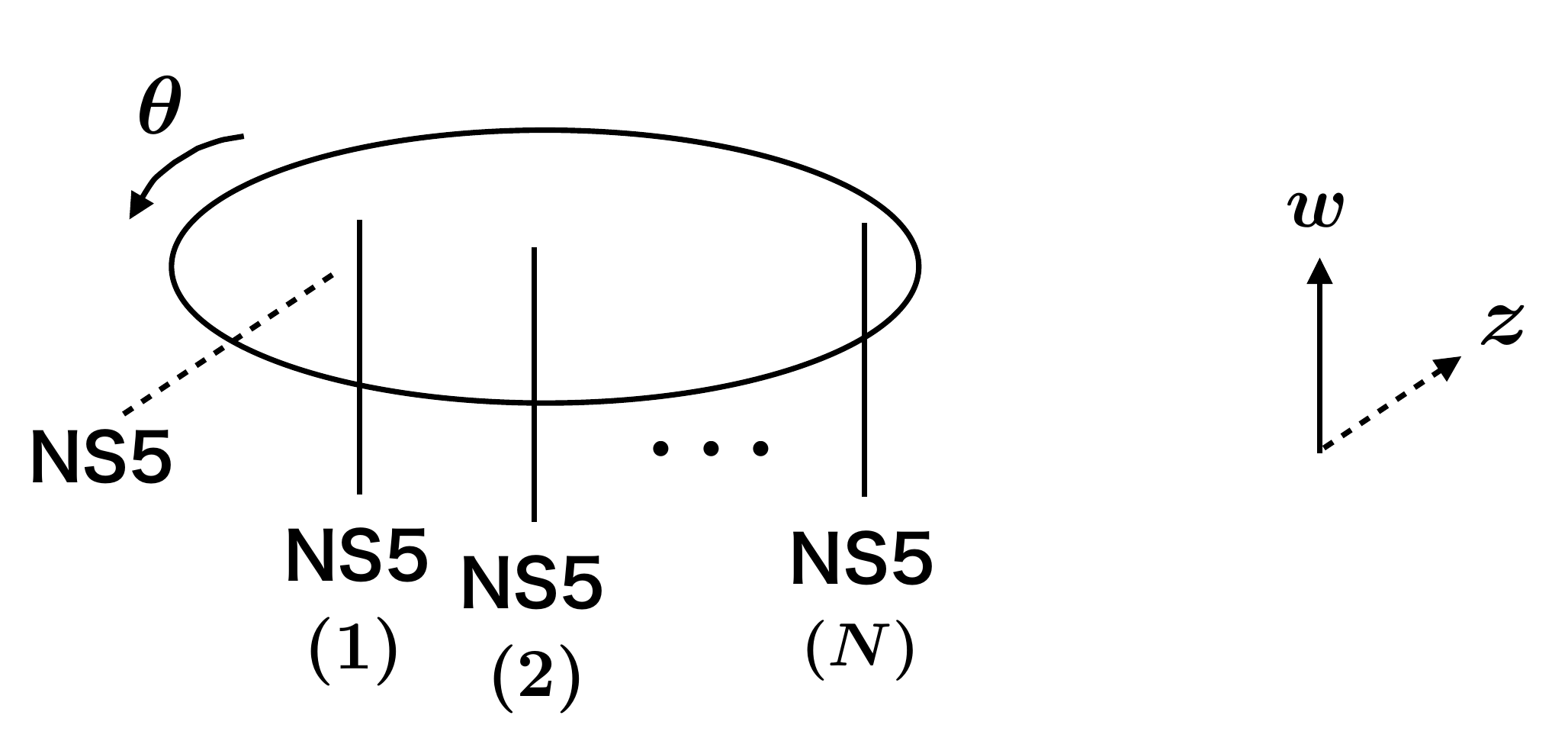}
\caption{The brane configuration in the type IIB side. There are $N$ parallel NS5 and one additional NS5'. The NS5-branes extend in $w$-plane, while the NS5' is localized in $w$-plane and extends in $z$-plane. The $\theta$-direction corresponds to the T-dual circle. Fractional D0-branes in the original IIA side are mapped to open D1-branes stretched between NS5-branes. A regular D0-brane is mapped to a closed D1-brane which can be decomposed into $N+1$ open D1-branes.}
\label{fig:NS5}
\end{center}
\end{figure}

In our case, since we have added an additional two-cycle $\beta_0$ to $A_{N-1}$ ALE space, a regular D0-brane is decomposed into $N+1$ fractional D0-branes, which is consistent with the feature of our model (figure \ref{fig:crystal}).  We can see this fact more explicitly in type IIB side. In this paper, we take the moduli parameters so that our Calabi-Yau geometry becomes singular. The singularity is expressed by a hypersurface
\begin{eqnarray}
 xy = z^Nw
\end{eqnarray}
in $\mathbb{C}^4$. By taking the T-duality transformation along the $U(1)$-orbit $x\to e^{i\theta}x,\,y\to e^{-i\theta}y$, this Calabi-Yau geometry is mapped to NS5-branes located on $x=y=0$ \cite{Uranga:1998vf}. The dual IIB side involves $N$ parallel NS5-branes which extend in $w$-plane (and located at $z=0$), and a single NS5-brane extending in $z$-plane (and located at $w=0$).\footnote{The NS5-branes also extend in the four-dimensional spacetime $R^{3,1}$ which is, in the original type IIA side, transverse to the Calabi-Yau geometry.} The brane configurations in $\theta$-direction can be depicted as in figure \ref{fig:NS5}.
A regular D0-brane in the original IIA side is now mapped to a closed D1-brane extending along $\theta$-direction, while the fractional D0-branes are mapped to open D1-branes stretched between two NS5-branes. From the figure \ref{fig:NS5}, we now find that a closed D1-brane can be divided into $N+1$ open D1-branes. This implies that on the original Calabi-Yau singularity a regular D0-brane can be decomposed into $N+1$ fractional D0-branes, which is consistent with the feature of our model.

\subsubsection*{Discussions}

In this article, we have constructed a statistical model  counting  the BPS D4-D2-D0 
bound states on a class of Calabi-Yau three-folds, in a small compact two-cycles region. The toric webdiagram of our Calabi-Yau is depicted in figure \ref{fig:diagram}. We have also found that our model is consistent with the relation between D2-branes on vanishing cycles and fractional D0-branes.

We should here mention the relation to the conifold case studied in \cite{Nishinaka:2011sv}. When we set $N=1$ in figure \ref{fig:diagram}, our Calabi-Yau three-fold becomes resolved conifold, and therefore the analysis in this article should reduce to that in \cite{Nishinaka:2011sv}. In fact, by setting $N=1$, our oblique partition model correctly reduces to ``the triangular partition model'' which was proposed in \cite{Nishinaka:2011sv} as a statistical model describing D4-D2-D0 states on the conifold. The triangular partition model involves red and blue triangle atoms but no parallelogram atom, and the rules of removing triangle atoms from the crystal is exactly the same as ours.

As a subject for future work, it would be interesting to study the relation between our model and the orbifold partitions \cite{Dijkgraaf:2007fe} describing the instantons on ALE space. In \cite{Nishinaka:2011nn}, the relation between our D4-D2-D0 states and the ALE instantons is studied. In particular, it was shown that the instanton partition function on $A_{N-1}$ ALE space is obtained by taking the {\em large radii limit} and neglecting D2-branes wrapping on the additional two-cycle $\beta_0$. On the other hand, our partition function \eqref{eq:partition_function} is valid in the {\em small radii limit.} It would be interesting to study the relation between the two statistical models.\footnote{When we change the moduli, the wall-crossing phenomena of D-branes occur. For the recent developments of D4-D2-D0 wall-crossings, see \cite{Diaconescu:2007bf, Jafferis:2007ti, Andriyash:2008it, Manschot:2009ia, Szabo:2009vw, Manschot:2010xp, Nishinaka:2010qk, Nishinaka:2010fh, Alim:2010cf, Nishinaka:2011sv, Nishinaka:2011pd}.}

It is also worth studying the relation to the free fermion description and the matrix model for BPS D-branes on a Calabi-Yau three-fold. At least for BPS D6-D2-D0 states, the statistical model description of the BPS D-branes is known to be closely related to free fermions and matrix model \cite{Ooguri:2010yk, Sulkowski:2010eg}.

Another interesting future direction will be to study the thermodynamic limit of our model. In \cite{Okounkov:2003sp, Iqbal:2003ds, Ooguri:2009ri}, it is pointed out that the thermodynamic limit of the crystal melting model for D6-D2-D0 states gives precise information of the smooth geometry of the mirror Calabi-Yau three-fold. It is worth studying the thermodynamic limit of our oblique partition model.


\section*{Acknowledgments}

We would like to thank Satoshi Yamaguchi for many illuminating discussions, important comments and suggestions.
We also thank Yoshifumi Hyakutake, Yosuke Imamura, Hiroaki Nakajima, Muneto Nitta, Kazutoshi Ohta and Tadashi Okazaki for useful discussions. The authors thank the Yukawa Institute for Theoretical Physics at Kyoto University, where this work was initiated and achieved during the YITP-W-11-05 on ``Development of Quantum Field Theory and String Theory.''


\bibliography{ref}

\end{document}